\documentclass[aps,prb,twocolumn]{revtex4}

\usepackage{graphicx}
\usepackage{dcolumn}
\usepackage{bm}
\usepackage{float}
\usepackage{amsfonts}
\usepackage{epsfig}

\begin{document}

\preprint{APS/XYZ}

\title{Structure and stability of the Si(331)-(12$\times$1) surface reconstruction}

\author{Corsin Battaglia}
\affiliation{Institut de Physique, Universit\'e de Neuch\^atel,
2000 Neuch\^atel, Switzerland}%
\email{corsin.battaglia@unine.ch}
\homepage{http://www.unine.ch/phys/spectro}
\author{Katalin Ga\'{a}l-Nagy}
\affiliation{Dipartimento di Fisica and European Theoretical
Spectroscopy Facility (ETSF), Universit\`a di Milano, 20133 Milano,
Italy}
\author{Giovanni Onida}
\affiliation{Dipartimento di Fisica and European Theoretical
Spectroscopy Facility (ETSF), Universit\`a di Milano, 20133 Milano,
Italy}
\author{Philipp Aebi}
\affiliation{Institut de Physique, Universit\'e de Neuch\^atel,
2000 Neuch\^atel, Switzerland}%

\date{\today}

\begin{abstract}
We recently proposed a new structural model for the
Si(331)-(12$\times$1) surface reconstruction containing silicon
pentamers as elementary structural building blocks. Using
first-principles density-functional theory we here investigate the
stability of a variety of adatom configurations and determine the
lowest energy configuration. We also present a detailed comparison
of the energetics between our model for Si(331)-(12$\times$1) and
the adatom-tetramer-interstitial model for Si(110)-(16$\times$2),
which shares the same structural building blocks.
\end{abstract}

 \keywords{}

\maketitle

\section{Introduction}

High-index silicon surfaces are of interest not only from a
fundamental point of view, but also as potential substrates for
electronic device fabrication. The reconstructed
Si(331)-(12$\times$1) surface is of particular importance since it
is the only confirmed planar silicon surface with a stable
reconstruction located between the (111) and (110) directions. Due
to its large unit cell and its pronounced structural anisotropy it
represents a promising template for the growth of low-dimensional
heteroepitaxial nanostructures such as quantum dots and nanowires. \cite{Battaglia08c}\\
Since its discovery more than 17 years ago \cite{Wei91} several
structural models containing dimers and adatoms as elementary
structural building blocks have been proposed.
\cite{Olshanetsky98,Gai01} We recently revealed the presence of an
additional building block on the Si(331)-(12$\times$1) surface.
\cite{Battaglia08} Using scanning tunneling microscopy (STM) we were
able to resolve for the first time rows of pentagon pairs running
across the Si(331) surface.
Very similar pentagons were observed earlier on the reconstructed
Si(110) and Ge(110) surfaces. \cite{An00,Ichikawa95,Gai98} Inspired
by structural elements encountered on Si(113) and Ge(113) surfaces,
\cite{Dabrowski95,Laracuente98,Stekolnikov03} An \textit{et al.}
\cite{An00} have proposed an adatom-tetramer-interstitial (ATI)
model for the Si(110)-(16$\times$2) reconstruction containing
silicon pentamers as building blocks which explain the pentagons
observed in STM images. The stability of the ATI model has
subsequently been tested theoretically by means of first-principles
total energy calculations. \cite{Stekolnikov04,Stekolnikov04b}
 Based on a detailed
analysis of our experimental results and a comparison between the
 Si(110) and Si(331) surface, we proposed a new
structural model for the Si(331)-(12$\times$1) reconstruction
containing silicon pentamers as essential building blocks.\cite{Battaglia08} \\
In order to account for the pentagons observed in our STM images,
the model contains two pentamers per (12$\times$1) unit cell, which
saturate a certain number of energetically unfavorable dangling
bonds of the bulk-truncated Si(331) surface. Some of the remaining
dangling bonds are saturated by simple adatoms. The pentamer
positions and bonding configurations were determined exclusively
using experimental information and by comparing Si(331) to Si(110)
as discussed in detail in Ref.~\onlinecite{Battaglia08}. Although
the available body of experimental results allows to narrow down the
location of the adatoms, currently we do not have sufficient
evidence to precisely determine the adatom positions on the surface
solely based on information derived from experiment. In this paper
we use first-principles total energy calculations to determine the
precise adatom positions by finding the lowest energy configuration
among various alternative adatom configurations.\\
The paper is organized as follows: After briefly reviewing the
methods used for the calculations, we focus on the bulk-truncated
Si(331) surface and compare its energetics to other silicon
surfaces, which are known to be stabilized by surface
reconstructions. In a second step pentamers are introduced and the
consequences on the total energy are discussed. Then we consider a
variety of candidate adatom configurations and determine the one
with the lowest energy. At the end we discuss the impact of the
remaining dangling bonds on the total energy.  \\

\begin{table*}[ht]
\caption{\label{tab:SurfaceEnergiesBT}Surface energies for unrelaxed
bulk-truncated surfaces of various orientations. The numbers in
parenthesis give the number of surface atoms and dangling bonds per
(1$\times$1) unit cell respectively.}
\begin{ruledtabular}
\begin{tabular}{rrrrr}
Surface & Surface atom & Dangling bond &   Estimated surface & Calculated surface\\
 orientation &    density (\AA$^{-2}$) & density (\AA$^{-2}$)&energy (meV/\AA$^2$) & energy (meV/\AA$^2$) \\
\hline
(111) &0.078 (1)&0.078 (1)&  90.1&113.6\cite{Stekolnikov02} \\
(100) &0.068 (1)&0.136 (2)& 155.9&149.2\cite{Stekolnikov02} \\
(113) &0.082 (2)&0.123 (3)& 141.0&137.9\cite{Stekolnikov03b} \\
(110) &0.096 (2)&0.096 (2)&  110.3&127.3\cite{Stekolnikov02} \\
(331) &0.093 (3)&0.093 (3)&  107.3&127.2     \footnotemark[1]
\end{tabular}
\end{ruledtabular}
\footnotetext[1]{this work}
\end{table*}

\section{Theory}
\subsection{Computational details}

Our results are based on first-principles calculations within
density-functional theory (DFT), \cite{Hoh64,Koh65} using the
local-density approximation (LDA) for the exchange-correlation
functional. \cite{Per81,Cep80} We adopt a plane-wave basis set and
norm-conserving pseudopotentials, as implemented in the PWSCF code
of the QUANTUM-ESPRESSO suite. \cite{PWscf}  Pseudopotentials for
silicon
and hydrogen are choosen in the von Barth style. \cite{CarUn,Cor93}\\
The Si(331) surface was simulated by means of an atomic slab and a
repeated supercell, in order to recover a three-dimensional
periodicity. The slab thickness corresponds to 10 silicon double
layers (DLs), and the periodic images are separated by 9.4 \AA$\;$
of vacuum. Since each bulk DL contains 24 silicon atoms, our
supercell counts 240 silicon atoms plus a number of silicon atoms at
the top surface (varying from 0 to 14 depending on the specific
reconstruction). The bottom surface of the slab is
hydrogen-passivated (with 36 hydrogen atoms).\\
Since we are interested in small energy differences, we adopt strict
convergence criteria with respect to both the kinetic-energy cutoff
and the sampling of the surface Brillouin zone. In particular, the
inclusion of plane waves up to 544 eV (40 Ry) and the use of a $xy$
shifted 3$\times$4$\times$1 Monkhorst-Pack grid \cite{Mon76} for the
{\bf k} point sampling allowed us to ideally compute total energy
differences with
 an accuracy of the order of 15 meV/supercell. By taking into account
 the fact that our relaxed structures still show small residual forces
 (see below),  we estimate our numerical accuracy on calculated surface
 energies to be better than  0.1 meV/\AA$^2$. All structures were optimized till the largest
residual force on the
mobile atoms was less than 0.026 eV/\AA$\;$ (0.001 Ry/Bohr).\\
The hydrogen atom positions were determined in a first run by
optimizing them on a bulk-terminated slab with fixed bulk-like
positions for silicon atoms, with top and bottom surfaces saturated
by hydrogen. In all the following calculations we have taken
hydrogen atomic positions at the bottom surface as fixed. In a
second step, we inspected the decay of forces in the direction
perpendicular to the surface, for a slab containing a non-relaxed
surface reconstruction. A nearly force-free bulk region is reached
after the five topmost DLs. Thus, only the topmost five bulk silicon
DL were allowed to relax together with the surface (reconstruction)
atoms, where the inner DLs have been kept fixed. All results
presented below are based on a slab relaxed in this way.

\subsection{Determination of the surface energy}

The surface energy $E_{\rm surf}$  of a double-sided symmetric slab
containing $N_{\rm Si}$ silicon atoms can be written as
\cite{Stekolnikov02}
\begin{eqnarray}
  E_{\rm surf} = \frac{1}{2} (E_{\rm tot}(N_{\rm Si})-\mu_{\rm Si} N_{\rm Si}) \quad ,
\end{eqnarray}
where $\mu_{\rm Si}$ is the chemical potential of silicon, being
simply the bulk energy per atom at zero temperature. This value can
be obtained from a simple bulk silicon calculation. $E_{\rm
tot}(N_{\rm Si})$ is the total energy of the slab. The factor of 2
comes from the fact that there are two identical surfaces. For a
non-symmetric slab for which only the bottom surface is saturated by
hydrogen atoms, we have
\begin{eqnarray}
  E_{\rm surf} + E^{\rm H}_{\rm surf} =
E_{\rm tot}(N_{\rm Si}, N_{\rm H})-\mu_{\rm Si} N_{\rm Si} -
\mu_{\rm H} N_{\rm H}
\end{eqnarray}
Here, $E^{\rm H}_{\rm surf}$ is the surface energy of the hydrogen
saturated surface, $\mu_{\rm H}$ is the chemical potential for the
hydrogen atoms, and $E_{\rm tot}(N_{\rm Si}, N_{\rm H})$ is the
total energy of the slab whose bottom surface is saturated by
$N_{\rm H}$ hydrogen atoms. Therefore, we have
\begin{eqnarray}
  E_{\rm surf} =
  E_{\rm tot}(N_{\rm Si}, N_{\rm H})-\mu_{\rm Si} N_{\rm Si} - \mu_{\rm H} N_{\rm H}
  - E^{\rm H}_{\rm surf} \quad .
\end{eqnarray}

In order to avoid the determination of $\mu_{\rm H}$, we consider
the following: The total energy of the slab can be written as
\begin{eqnarray}
  E_{\rm tot}(N_{\rm Si}, N_{\rm H}) = \frac{1}{2} E_{\rm tot}^{\rm H}
              + \frac{1}{2} E_{\rm tot}
\end{eqnarray}
where $E_{\rm tot}^{\rm H}$ is the total energy of a symmetric
hydrogen saturated slab and $E_{\rm tot}$ is the total energy of a
symmetric slab containing the surface reconstruction. The value
$E_{\rm tot}^{\rm H}$ has been calculated in the preliminary run
used for the determination of the hydrogen positions. The surface
energy (per unit area) for the top surface carrying the surface
reconstruction can then be determined using
\begin{eqnarray}
\gamma=  E_{\rm surf}/A
  =\big( E_{\rm tot}(N_{\rm Si}, N_{\rm H})  - \frac{1}{2} E_{\rm tot}^{\rm H}
    - \tilde{N}_{\rm Si}\mu_{\rm Si}\big)/A
\end{eqnarray}
where $\tilde{N}_{\rm Si}$ is the number of silicon atoms in the
upper half slab and $A$ the surface area of the slab.

\section{Results and discussion}

\begin{figure}[b]
\centering
\includegraphics{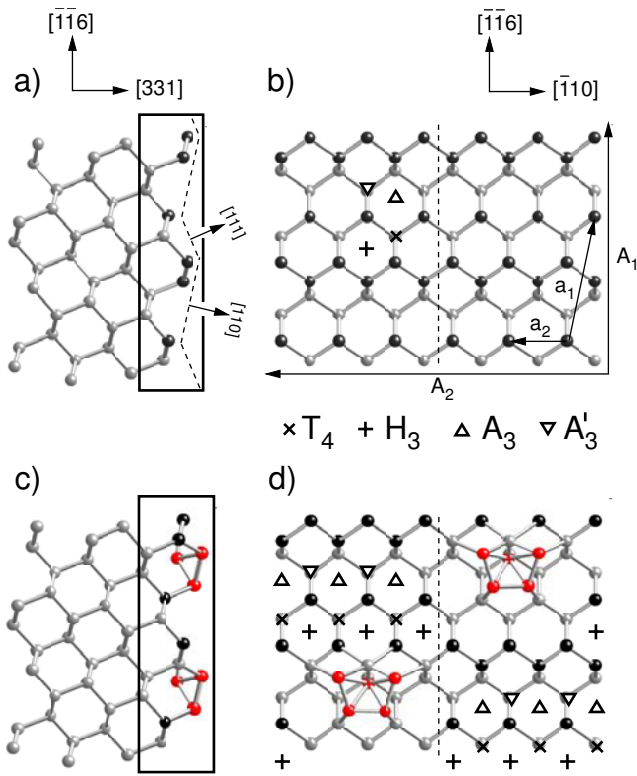}%
\caption{\label{Fig1}(Color online) Side (a)/(c) and top view
(b)/(d) of the unrelaxed bulk-truncated Si(331)-(12$\times$1)
surface unit cell/unrelaxed 'pentamers only' model. The bulk
directions are given. In (b)/(d) only the topmost atoms contained
within the rectangle in (a)/(c) are shown. Subsurface atoms are
shown in grey, undercoordinated surface atoms in black, pentamer
atoms in red. In (a) the Si(111) and Si(110) terraces composing the
Si(331) surface are indicated. In (b) the (1$\times$1) and
(12$\times$1) surface unit cell vectors $\mathbf{a_1}$,
$\mathbf{a_2}$ and $\mathbf{A_1}$, $\mathbf{A_2}$ respectively are
also drawn. The orientation of the glide plane is indicated by a
dashed line. In (b) the four different types of adatom adsorption
sites $T_4$, $H_3$, $A_3$ and $A'_3$ are represented by markers. In
(d) the 2$\times$12 possible adatom positions after introduction of
the two pentamers are shown.}
\end{figure}

\subsection{Bulk-truncated surface}

In order to define an energy reference for the following discussion
we have determined the surface energy for the unrelaxed
bulk-truncated Si(331) surface shown in Fig.~\ref{Fig1}(a) and (b).
The value of 127.2 meV/\AA$^2$ may be compared in
Tab.~\ref{tab:SurfaceEnergiesBT} with the values for
other surface orientations.\\
The surface energy of a bulk-truncated surface can be estimated from
the cohesive energy of silicon, which is $E_c=4.6$ eV per atom.
\cite{Farid91} The cohesive energy of a solid is the energy per atom
required to break the atoms of the solid into isolated atomic
species. Since a bond is formed by two atoms and each silicon atom
makes four bonds, the energy required to break a bond is half the
cohesive energy, i.e. $E_b=2.3$ eV per bond. \cite{Harrison80} The
surface energy $\gamma$ can now be estimated by multiplying the bond
energy $E_b$ by the density of broken bonds $n$. \cite{Dabrowski00}
The latter is the number of dangling bonds $N$ per surface area $A$,
$n=N/A$. Thus $\gamma= N E_b/2A= n E_b/2$. The factor of 2 takes
into account that two
surfaces are created.\\
Note that the dangling bond density $n$ is not necessarily equal to
the surface atom density (see values in parenthesis in
Tab.~\ref{tab:SurfaceEnergiesBT}). This is due to the fact that a
surface atom may either carry one or two dangling bonds. (111)-like
surface atoms carry a single dangling bond, whereas (100)-like
surface atoms carry two dangling bonds (see Fig. 3(a) and 2(a)
respectively in Ref.~\onlinecite{Battaglia08b}). On the (113)
surface, both types of surface atoms coexists, whereas on the (110)
and the (331) surface only (111)-type surface atoms occur (see Fig.
5(a), 6(a) and 7(a)
respectively in Ref.~\onlinecite{Battaglia08b}).\\
In Tab.~\ref{tab:SurfaceEnergiesBT} we summarize estimated and
calculated surface energies of unrelaxed bulk-truncated surfaces for
a series of surface orientations. Although the simple estimate based
on the cohesive energy fails to precisely reproduce the values
determined with the help of first-principles methods, it correctly
reproduces the energy ordering of the various surfaces. The
estimated values tend to underestimate the surface energy for
surfaces containing only (111)-type surface atoms, whereas they
overestimate the surface energy for surfaces with (100)-type surface
atoms. For Si(113), which contains both types of surface atoms,
these two tendencies compensate resulting in the best agreement
between estimated and
calculated value.\\
From the theoretical point of view, the most stable surface is the
one with the lowest surface energy. Consequently only structural
models for the reconstructed surface with a surface energy lower
than the bulk-truncated surface must be considered. Allowing the
bulk-truncated surface to relax and determining the corresponding
surface energy, we obtain an even more stringent condition on the
surface energy for any valid candidate model. The relaxed structure
of the bulk-truncated Si(331) surface is shown in
Fig.~\ref{Fig2}(a). Note that although the relaxation was performed
within the (12$\times$1) supercell, it did not break the
(1$\times$1) periodicity within our numerical accuracy. The surface
energy for the relaxed bulk-truncated surface is given in
Tab.~\ref{tab:SurfaceEnergiesComp} along with the surface energies
for other structures for the Si(331) and Si(110) surface discussed
hereafter. Relaxation of the bulk-truncated Si(331) surface within
the (12$\times$1) unit cell results in a reduction of the surface
energy of 16.3 meV/\AA$^2$, a value to be compared with the
corresponding value of 21.2 meV/\AA$^2$ for Si(110) obtained by
Stekolnikov \textit{et al.} \cite{Stekolnikov04b} via relaxation of
the bulk-truncated Si(110) surface within the (16$\times$2) cell.

\subsection{Pentamers}

\begin{figure}[h!]
\centering
\includegraphics{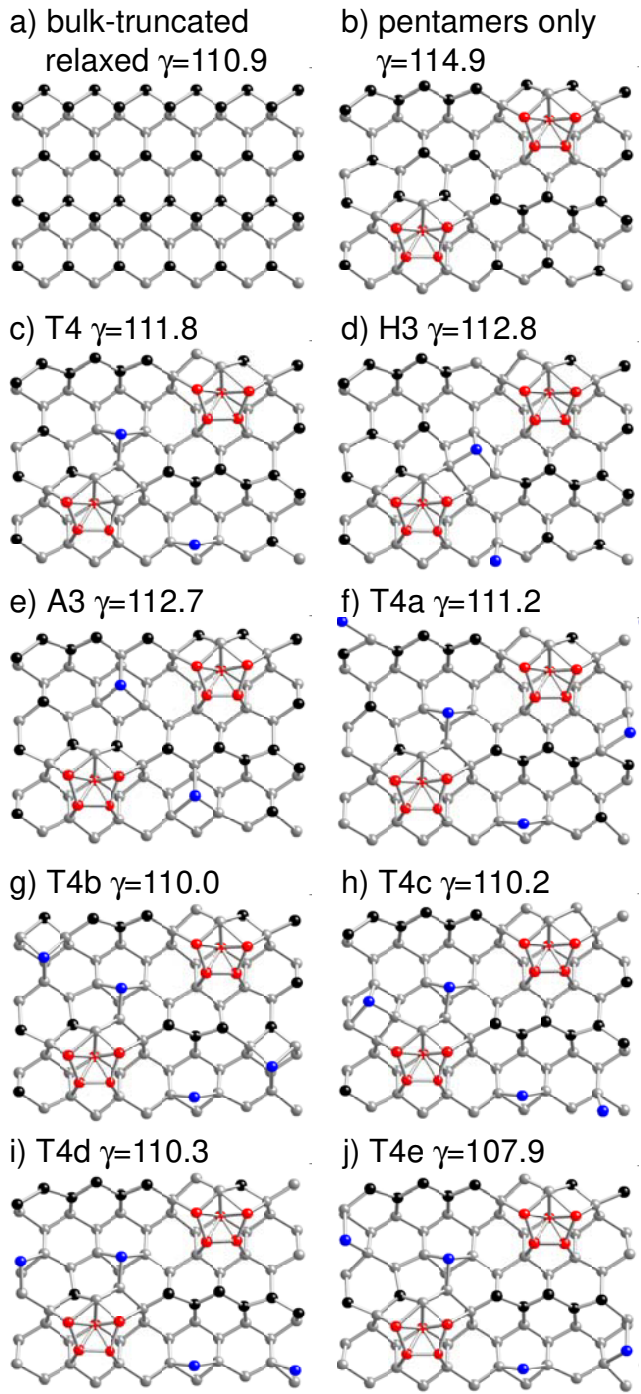}%
\caption{\label{Fig2}(Color online) (a) Top view of the relaxed
bulk-truncated Si(331)-(12$\times$1) surface. (b) Top view of the
relaxed 'pentamers only' model. (c)-(j) Various candidate adatom
configurations of the adatom-pentamer model. The surface energy
$\gamma$ for each model is given in units of meV/\AA$^2$. }
\end{figure}

\begin{table}[t]
\caption{\label{tab:SurfaceEnergiesComp}Comparison between the
surface energies of various surface structures of Si(331) and
Si(110) }
\begin{ruledtabular}
\begin{tabular}{lll}
        &  \multicolumn{2}{l}{Surface energy (meV/\AA$^2$)} \\
Structure  & Si(331)& Si(110) \\
 \hline
bulk-truncated   & 127.2\footnotemark[1]  & 127.3\cite{Stekolnikov02} \\
bulk-truncated relaxed    & 110.9\footnotemark[1] & 106.1\cite{Stekolnikov04b,Stekolnikov02} \\
pentamers only   &  114.9\footnotemark[1]  & 108.6\cite{Stekolnikov04b,Stekolnikov02}\\
adatoms and pentamers   & 107.9\footnotemark[1]&
103.8\cite{Stekolnikov04b,Stekolnikov02}
\end{tabular}
\end{ruledtabular}
   \footnotetext[1]{this work}
\end{table}

The bulk-truncated Si(331) surface consists of alternating single
and double rows of undercoordinated surface atoms running along the
$[\bar{1}10]$ direction (black atoms in Fig.~\ref{Fig1}(a) and (b)).
Each of the 36 (111)-type surface atoms per (12$\times$1) unit cell
carries one dangling bond, which
renders the unreconstructed structure energetically unfavorable.\\
In our model, 10 out of the 36 dangling bonds are saturated by two
silicon pentamers observed as pentagons in STM images. Their
integration onto the bulk-truncated Si(331) surface was discussed in
detail in Ref.~\onlinecite{Battaglia08}. It is important to note
that the bulk-truncated Si(331) surface may be viewed as a highly
stepped surface consisting of small alternating Si(111) and Si(110)
terraces (see Fig.~\ref{Fig1}(a)). This allows to anchor the
pentamers on the Si(331) surface in exactly the same local binding
configuration as on the Si(110) surface, where pentamers are also
observed. Introducing two pentamers per unit cell as shown in
Fig.~\ref{Fig1}(c) and (d) reduces the number of dangling bonds from
36 to 26 and lowers the surface energy (for the corresponding
relaxed model shown in Fig.~\ref{Fig2}(b)) by 12.3 meV/\AA$^2$ with
respect to the unrelaxed bulk-truncated surface (see
Tab.~\ref{tab:SurfaceEnergiesComp}). This value is comparable with
the reduction of 18.7 meV/\AA$^2$ obtained when introducing four
pentamers onto the Si(110)-(16$\times$2) surface (corresponding to
the tetramer-interstitial (TI) model in
Ref.~\onlinecite{Stekolnikov04b}, note that in the TI model the step
is already included into the unit cell accounting for 1.0
meV/\AA$^2$ energy reduction). However, the 'pentamers only' models
for both, Si(331) and Si(110), are not stable structures, since the
relaxed bulk-truncated surface for both surfaces possesses a lower
surface energy. This is not surprising, since the introduction of
pentamers suggested by STM images, leaves a lot of dangling bonds
unsaturated. So an additional building block is required to
stabilize the structures containing pentamers.

\subsection{Adatoms}

In the ATI model for Si(110)-(16$\times$2), the pentamer structure
is stabilized by adatoms. Among the various elementary structural
building blocks observed in silicon surface reconstructions (see
Ref.~\onlinecite{Battaglia08b} for a review), adatoms impose
themselves also as the best choice for Si(331) to saturate some of
the remaining 26 dangling bonds for the following two reasons:
Firstly, as mentioned before, the bulk-truncated Si(331) surface may
be viewed as being composed of small alternating (110) and (111)
terraces. Since adatoms are observed on both the Si(110) and the
Si(111) surface, the Si(331) consequently provides exactly the same
local environment for adatoms as on Si(110) and Si(111). Secondly
adatoms allow to explain the additional protrusions seen in STM
images, which can not be attributed to the
pentamer atoms. \cite{Battaglia08}\\
An adatom requires three unsaturated surface atoms to be attached
to. On the (111) terraces of the Si(331) surface, adatoms may occupy
two possible sites indicated in Fig.~\ref{Fig1}(b). These geometries
are distinguished as hollow ($H_3$) and atop ($T_4$) sites depending
on whether the substrate atoms below the adatom is found in the
fourth or second layer. In $H_3$ sites the adatom is three-fold
coordinated, in $T_4$ sites the adatom is approximately four-fold
coordinated due to the substrate atom directly below in the second
layer.\\
On the (110) terraces of the Si(331) surface, we can also
distinguish between two different three-fold coordinated adsorption
sites depending on their orientation labeled $A_3$ and $A'_3$ in
Fig.~\ref{Fig1}(b). Note that on the bulk-truncated (110) surface
these two sites become equivalent due to the mirror symmetry
of the (110) surface along the $[\bar{1}10]$ direction (not shown).\\
Careful inspection of Fig.~\ref{Fig1}(d) shows that after
introducing two pentamers per unit cell there are exactly
2$\times$12 possible positions for placing adatoms, the factor of 2
is due to the glide plane symmetry along the $[\bar{1}\bar{1}{6}]$
direction (dashed line in Fig.~\ref{Fig1}(d)). Out of the
2$\times$12 candidate adsorption sites, we identify 2$\times$3 $T_4$
sites, 2$\times$4 $H_3$ sites, 2$\times$3 $A_3$ sites and 2$\times$2
$A'_3$ sites. This however does not imply that a maximum of 24
adatoms can be placed on the surface. The reason for this is as
follows. After placing a first adatom on the surface, some
surrounding candidate sites can not be occupied by an adatom
anymore, since they shared the same dangling bond with the now
occupied site, whose
adatom saturates this dangling bond.\\
A first pair of adatoms must be placed between two successive
pentamers, in order to account for the clear protrusions seen in STM
images. \cite{Battaglia08} Due to the glide plane symmetry adatoms
must always occur in pairs. There are only three candidate sites for
this first adatom: a $T_4$ site (Fig.~\ref{Fig2}(c)), a $H_3$ site
(Fig.~\ref{Fig2}(d)) and a $A_3$ site (Fig.~\ref{Fig2}(e)). Using
first-principles calculations we have tested these three possible
adatom configurations and found that the $T_4$ site is the most
stable of the three (surface energies for each model are also given
in Fig.~\ref{Fig2}). This is not surprising, since $T_4$ sites are
commonly favored over the $H_3$ sites on (111) surfaces such as in
the Si(111)-(7$\times$7) reconstruction and the
Ge(111)-c(2$\times$8) reconstruction (see
Ref.~\onlinecite{Battaglia08b} for a review of their structural
models). In addition the $T_4$ adatom position reproduces best the
protrusion observed by STM between two successive pentagons. \cite{Battaglia08}\\
Consequently, in order to reduce the large number of potential
candidate adatom configurations to a computationally manageable
number of models, we have restricted our further investigations to
the subset of models containing the $T_4$ adatom pair. In a second
step we have investigated \emph{all} possible remaining positions
for a second pair of adatoms. There are exactly five candidate
configurations to be considered (shown in Fig.~\ref{Fig2}(f)-(j)).
Using first-principles calculations we find that a $A_3$ adatom
position (configuration T4e) gives the most stable configuration.
The position of the $A_3$ adatom pair is also in agreement with
light protrusions seen in STM images \cite{Battaglia08}. After
placing this $A_3$ adatom pair there is no more
candidate site left on which one could place an additional adatom.\\
As discussed before, the introduction of two pentamers is not
sufficient to obtain a lower surface energy than for the relaxed
bulk-truncated surface. Inspection of
Tab.~\ref{tab:SurfaceEnergiesComp} shows that adding adatoms to the
pentamers allows to sufficiently lower the surface energy to render
the reconstructed surface more stable than the relaxed
bulk-truncated surface, for both the Si(331) and the Si(110)
surfaces (reduction of surface energy by 7.0 meV/\AA$^2$ and 4.8
meV/\AA$^2$ respectively with respect to the 'pentamers only'
models).

\begin{table}[t]
\caption{\label{tab:SurfaceEnergiesRec}Surface energies for silicon
surface reconstructions }
\begin{ruledtabular}
\begin{tabular}{lc}
Reconstruction &  Surface energy  \\
   & (meV/\AA$^2$) \\
\hline
Si(111)-(7$\times$7) &  84.9\cite{Stekolnikov02}                     \\
Si(100)-c(4$\times$2)&  88.0\cite{Stekolnikov02}                      \\
Si(113)-(3$\times$2) &
87.9\cite{GaalNagy07}/87.4\cite{Stekolnikov02}  \\
Si(110)-(16$\times$2)& 103.8\cite{Stekolnikov04b,Stekolnikov02}                      \\
Si(331)-(12$\times$1)& 106.4\footnotemark[1] \\
\end{tabular}
\end{ruledtabular}
   \footnotetext[1]{this work}
\end{table}

\subsection{Rest atoms}

Introducing two $T_4$ and two $A_3$ adatoms per unit cell into our
structural model for the Si(331)-(12$\times$1) reconstruction
further reduces the number of dangling bonds from 26 to 14 (10 of
the 36 dangling bonds are saturated by the pentamers, introducing 4
adatoms further eliminates 4$\times$3 dangling bonds (here we do not
count the dangling bond carried by each adatom)). The remaining 14
surface atoms, which still carry a
dangling bond, are so called rest atoms (black atoms in Fig. \ref{Fig2}(i)).\\
Rest atoms are common structural building blocks accompanying
adatoms. The adatom model for the Ge(111)-c(2$\times$8)
reconstruction is built from four adatoms and four rest atoms. For
this case, we have an ideal one-to-one correspondence between
adatoms and rest atoms. Each adatom delivers one electron from its
dangling bond to fill the dangling bond of the rest atom resulting
in a further decrease of the surface energy. In contrast, in the
famous dimer-adatom-stacking fault (DAS) model of the
Si(111)-(7$\times$7) reconstruction, there are 12 adatoms, but only
6 rest atoms plus the corner hole atom, which behaves similarly to a
rest atom. In this case, using a simple ionic picture, we have 7
electrons which are transferred from the adatoms into the rest atom
plus corner hole atom states, but five electrons remain in the
adatoms bands. For the ATI model of the Si(110)-(16$\times$2)
reconstruction, there are 8 adatoms and 12 rest atoms. So even when
transferring the 8 electrons from the adatoms to the rest atoms,
only a fraction of the rest atoms may saturate their dangling bonds.
Similarly in our model, with four adatoms and 14 rest atoms, we
expect some remaining unsaturated rest atoms.\\
The unequal adatom:rest atom ratio between the ATI model (8:12=0.67)
of Si(110) and our model (4:14=0.29) of Si(331) might on the first
view explain why the ATI model exhibits a slightly lower surface
energy than our model (energy difference 4.1 meV/\AA$^2$, see
Tab.~\ref{tab:SurfaceEnergiesComp}). Reinspection of
Tab.~\ref{tab:SurfaceEnergiesComp} however reveals that this energy
difference between Si(331) and Si(110), whose corresponding
unrelaxed bulk-truncated surfaces have almost the same energy
(energy difference only 0.1 meV/\AA$^2$), is already present for the
respective 'pentamers only' models (energy difference between
'pentamers only' models 6.3 meV/\AA$^2$). It is also interesting to
note that the introduction of the adatoms leads to a larger
reduction of the surface energy for Si(331) (7.0 meV/\AA$^2$) than
for Si(110) (4.8 meV/\AA$^2$) with respect to the
'pentamers only' models.\\
Introducing hypothetically an additional adatom pair into our model
for Si(331) would improve its adatom:rest atom ratio from 4:14=0.29
to 6:8=0.75. The introduction of the first pair of adatoms ($T_4$)
accounts for a reduction of 3.1 meV/\AA$^2$ of the surface energy.
Adding a second pair of adatoms ($A_3$) to the first pair allowed to
reduce the surface energy by 3.9 meV/\AA$^2$. Following this
tendency, by assuming an energy gain of 3-4 meV/\AA$^2$ for the
introduction of a hypothetical third pair of adatoms into our model,
suggests that the surface energy of Si(331) comes to lie close to
the value determined for Si(110) by Stekolnikov \textit{et al.}
\cite{Stekolnikov04b}. However, note that the rest atom geometry of
T4e does not allow the introduction of an additional adatom pair.
When considering the subset of T4x models (with x=a-e), only
configurations T4a and T4b admit a pair of free adatom sites,
resulting in configuration T4a+b. But this configuration was
discarded because it is not compatible with the experimental
STM images. \cite{Battaglia08} \\
Several more three adatom pair configurations are possible when
studying H3x (8/7 possible two/three adatom pair configurations) or
A3x configurations (5/1 possible two/three adatom pair
configurations). Even more possibilities may exist when allowing the
placement of the first adatom pair on a different site. Here we
focused on the family of T4x models, whose selection was motivated
and justified in section C. Before testing other potential
configurations, further experimental hints on the position of the
adatoms from complementary experimental techniques are necessary to
assist in the search for the lowest energy configuration.
\\
When compared to the surface energies of reconstructions on surfaces
with other orientations listed in Tab. \ref{tab:SurfaceEnergiesRec},
the models for the Si(110)-(16$\times$2) and the
Si(331)-(12$\times$1) reconstruction have surface energies which are
of the order of 20 meV/\AA$^2$ higher than the surface energies for
Si(111)-(7$\times$7), Si(100)-c(4$\times$2) and
Si(113)-(3$\times$2). As it was already mentioned by Stekolnikov
\textit{et al.} \cite{Stekolnikov04b} in the context of the ATI
model for Si(110)-(16$\times$2), a completely new, more complex
reconstruction model cannot be excluded. Nevertheless, the advantage
of our method for developing a structural model for
Si(331)-(12$\times$1) consists here in integrating elementary
structural building blocks such as pentamers and adatoms identified
on the (110) and (111) silicon surfaces \cite{Battaglia08b} onto the
bulk-truncated Si(331) surface. The consistency of the pentamer
building block with all currently available experimental data on
Si(331)-(12$\times$1), but also on Si(110)-(16$\times$2), represents
further indirect evidence in favor of the two structural models. It
should be noted, that our model for Si(331) is far from a simple
extension of the Si(110) surface to its vicinals. The pentagons on
Si(110) (see Fig. 6 in Ref.~\onlinecite{Battaglia08b}) are arranged
in pairs pointing in opposite directions, whereas on Si(331) pairs
point in the same direction. Also the direction of the double
pentagon rows appears to be completely unrelated between the two
surfaces. Using pentamers and adatoms we succeeded in combining
these seemingly
conflicting experimental observations in a coherent picture.\\

\section{Summary and conclusions}

Using first-principles density-functional theory we studied the
stability of the structural model for the Si(331)-(12$\times$1)
surface reconstruction, which we recently proposed. In particular we
determined the energetically most favorable adatom configuration
among a variety of candidate structures. We find that a
configuration with a pair of $T_4$ adatoms and a second pair of
$A_3$ adatoms results in the lowest energy. We also discussed the
implications of the rest atoms on the energy and compared our model
to other silicon surface reconstructions.

\section*{Acknowledgments}

Stimulating discussions with Claude Monney, Cl\'{e}ment Didiot, Eike
Schwier, Nicolas Mariotti, and Michael Gunnar Garnier are gratefully
acknowledged. Skillfull technical assistance was provided by our
workshop and electric engineering team. This work was supported by
the Fonds National Suisse pour la Recherche Scientifique through
Div. II and the Swiss National Center of Competence in Research
MaNEP. We acknowledge support by the European Community through the
"Nanoquanta  NoE" and "ETSF-I3" projects (NMP4-CT-2004-500198 and
Grant  Agreement n. 211956) for the theoretical results, which are
an outcome of the ETSF User Project n. 94.

\bibliography{Si331_Theory}

\begin{thebibliography}{28}
\expandafter\ifx\csname natexlab\endcsname\relax\def\natexlab#1{#1}\fi
\expandafter\ifx\csname bibnamefont\endcsname\relax
  \def\bibnamefont#1{#1}\fi
\expandafter\ifx\csname bibfnamefont\endcsname\relax
  \def\bibfnamefont#1{#1}\fi
\expandafter\ifx\csname citenamefont\endcsname\relax
  \def\citenamefont#1{#1}\fi
\expandafter\ifx\csname url\endcsname\relax
  \def\url#1{\texttt{#1}}\fi
\expandafter\ifx\csname urlprefix\endcsname\relax\def\urlprefix{URL }\fi
\providecommand{\bibinfo}[2]{#2}
\providecommand{\eprint}[2][]{\url{#2}}

\bibitem[{\citenamefont{Battaglia
  et~al.}({\natexlab{a}})\citenamefont{Battaglia, Monney, Didiot, Schwier,
  Mariotti, Garnier, and Aebi}}]{Battaglia08c}
\bibinfo{author}{\bibfnamefont{C.}~\bibnamefont{Battaglia}},
  \bibinfo{author}{\bibfnamefont{C.}~\bibnamefont{Monney}},
  \bibinfo{author}{\bibfnamefont{C.}~\bibnamefont{Didiot}},
  \bibinfo{author}{\bibfnamefont{E.~F.} \bibnamefont{Schwier}},
  \bibinfo{author}{\bibfnamefont{M.}~\bibnamefont{Mariotti}},
  \bibinfo{author}{\bibfnamefont{M.~G.} \bibnamefont{Garnier}},
  \bibnamefont{and} \bibinfo{author}{\bibfnamefont{P.}~\bibnamefont{Aebi}},
  \bibinfo{note}{arxiv:0809.4967, submitted to AIP Conference Proceedings}.

\bibitem[{\citenamefont{Wei et~al.}(1991)\citenamefont{Wei, Williams, and
  Park}}]{Wei91}
\bibinfo{author}{\bibfnamefont{J.}~\bibnamefont{Wei}},
  \bibinfo{author}{\bibfnamefont{E.~D.} \bibnamefont{Williams}},
  \bibnamefont{and} \bibinfo{author}{\bibfnamefont{R.~L.} \bibnamefont{Park}},
  \bibinfo{journal}{Surf. Sci. Lett.} \textbf{\bibinfo{volume}{250}},
  \bibinfo{pages}{L368} (\bibinfo{year}{1991}).

\bibitem[{\citenamefont{Olshanetsky et~al.}(1998)\citenamefont{Olshanetsky,
  Teys, and Kozhemyako}}]{Olshanetsky98}
\bibinfo{author}{\bibfnamefont{B.~Z.} \bibnamefont{Olshanetsky}},
  \bibinfo{author}{\bibfnamefont{S.~A.} \bibnamefont{Teys}}, \bibnamefont{and}
  \bibinfo{author}{\bibfnamefont{I.~G.} \bibnamefont{Kozhemyako}},
  \bibinfo{journal}{Phys. Low-Dim. Struct.} \textbf{\bibinfo{volume}{11/12}},
  \bibinfo{pages}{85} (\bibinfo{year}{1998}).

\bibitem[{\citenamefont{Gai et~al.}(2001)\citenamefont{Gai, Zhao, Sakurai, and
  Yang}}]{Gai01}
\bibinfo{author}{\bibfnamefont{Z.}~\bibnamefont{Gai}},
  \bibinfo{author}{\bibfnamefont{R.~G.} \bibnamefont{Zhao}},
  \bibinfo{author}{\bibfnamefont{T.}~\bibnamefont{Sakurai}}, \bibnamefont{and}
  \bibinfo{author}{\bibfnamefont{W.~S.} \bibnamefont{Yang}},
  \bibinfo{journal}{Phys. Rev. B} \textbf{\bibinfo{volume}{63}},
  \bibinfo{pages}{085301} (\bibinfo{year}{2001}).

\bibitem[{\citenamefont{Battaglia
  et~al.}({\natexlab{b}})\citenamefont{Battaglia, Ga\'{a}l-Nagy, Monney,
  Didiot, Schwier, Garnier, Onida, and Aebi}}]{Battaglia08}
\bibinfo{author}{\bibfnamefont{C.}~\bibnamefont{Battaglia}},
  \bibinfo{author}{\bibfnamefont{K.}~\bibnamefont{Ga\'{a}l-Nagy}},
  \bibinfo{author}{\bibfnamefont{C.}~\bibnamefont{Monney}},
  \bibinfo{author}{\bibfnamefont{C.}~\bibnamefont{Didiot}},
  \bibinfo{author}{\bibfnamefont{E.~F.} \bibnamefont{Schwier}},
  \bibinfo{author}{\bibfnamefont{M.~G.} \bibnamefont{Garnier}},
  \bibinfo{author}{\bibfnamefont{G.}~\bibnamefont{Onida}}, \bibnamefont{and}
  \bibinfo{author}{\bibfnamefont{P.}~\bibnamefont{Aebi}},
  \bibinfo{note}{arxiv:0807.3875, accepted for publication in Phys. Rev. Lett.}

\bibitem[{\citenamefont{An et~al.}(2000)\citenamefont{An, Yoshimura, Ono, and
  Ueda}}]{An00}
\bibinfo{author}{\bibfnamefont{T.}~\bibnamefont{An}},
  \bibinfo{author}{\bibfnamefont{M.}~\bibnamefont{Yoshimura}},
  \bibinfo{author}{\bibfnamefont{I.}~\bibnamefont{Ono}}, \bibnamefont{and}
  \bibinfo{author}{\bibfnamefont{K.}~\bibnamefont{Ueda}},
  \bibinfo{journal}{Phys. Rev. B} \textbf{\bibinfo{volume}{61}},
  \bibinfo{pages}{3006} (\bibinfo{year}{2000}).

\bibitem[{\citenamefont{Ichikawa et~al.}(1995)\citenamefont{Ichikawa, Sueyosi,
  Saot, Iwatsuki, Udagwa, and Sumita}}]{Ichikawa95}
\bibinfo{author}{\bibfnamefont{T.}~\bibnamefont{Ichikawa}},
  \bibinfo{author}{\bibfnamefont{T.}~\bibnamefont{Sueyosi}},
  \bibinfo{author}{\bibfnamefont{T.}~\bibnamefont{Saot}},
  \bibinfo{author}{\bibfnamefont{M.}~\bibnamefont{Iwatsuki}},
  \bibinfo{author}{\bibfnamefont{F.}~\bibnamefont{Udagwa}}, \bibnamefont{and}
  \bibinfo{author}{\bibfnamefont{I.}~\bibnamefont{Sumita}},
  \bibinfo{journal}{Sol. Stat. Comm.} \textbf{\bibinfo{volume}{93}},
  \bibinfo{pages}{541} (\bibinfo{year}{1995}).

\bibitem[{\citenamefont{Gai et~al.}(1998)\citenamefont{Gai, Zhao, and
  Yang}}]{Gai98}
\bibinfo{author}{\bibfnamefont{Z.}~\bibnamefont{Gai}},
  \bibinfo{author}{\bibfnamefont{R.~G.} \bibnamefont{Zhao}}, \bibnamefont{and}
  \bibinfo{author}{\bibfnamefont{W.~S.} \bibnamefont{Yang}},
  \bibinfo{journal}{Phys. Rev. B} \textbf{\bibinfo{volume}{57}},
  \bibinfo{pages}{R6795} (\bibinfo{year}{1998}).

\bibitem[{\citenamefont{Dabrowski et~al.}(1994)\citenamefont{Dabrowski,
  M\"ussig, and Wolff}}]{Dabrowski95}
\bibinfo{author}{\bibfnamefont{J.}~\bibnamefont{Dabrowski}},
  \bibinfo{author}{\bibfnamefont{H.-J.} \bibnamefont{M\"ussig}},
  \bibnamefont{and} \bibinfo{author}{\bibfnamefont{G.}~\bibnamefont{Wolff}},
  \bibinfo{journal}{Phys. Rev. Lett.} \textbf{\bibinfo{volume}{73}},
  \bibinfo{pages}{1660} (\bibinfo{year}{1994}).

\bibitem[{\citenamefont{Laracuente et~al.}(1998)\citenamefont{Laracuente,
  Erwin, and Whitman}}]{Laracuente98}
\bibinfo{author}{\bibfnamefont{A.}~\bibnamefont{Laracuente}},
  \bibinfo{author}{\bibfnamefont{S.~C.} \bibnamefont{Erwin}}, \bibnamefont{and}
  \bibinfo{author}{\bibfnamefont{L.~J.} \bibnamefont{Whitman}},
  \bibinfo{journal}{Phys. Rev. Lett.} \textbf{\bibinfo{volume}{81}},
  \bibinfo{pages}{5177} (\bibinfo{year}{1998}).

\bibitem[{\citenamefont{Stekolnikov
  et~al.}(2003{\natexlab{a}})\citenamefont{Stekolnikov, Furthm\"uller, and
  Bechstedt}}]{Stekolnikov03}
\bibinfo{author}{\bibfnamefont{A.~A.} \bibnamefont{Stekolnikov}},
  \bibinfo{author}{\bibfnamefont{J.}~\bibnamefont{Furthm\"uller}},
  \bibnamefont{and}
  \bibinfo{author}{\bibfnamefont{F.}~\bibnamefont{Bechstedt}},
  \bibinfo{journal}{Phys. Rev. B} \textbf{\bibinfo{volume}{67}},
  \bibinfo{pages}{195332} (\bibinfo{year}{2003}{\natexlab{a}}).

\bibitem[{\citenamefont{Stekolnikov
  et~al.}(2004{\natexlab{a}})\citenamefont{Stekolnikov, Furthm\"uller, and
  Bechstedt}}]{Stekolnikov04}
\bibinfo{author}{\bibfnamefont{A.~A.} \bibnamefont{Stekolnikov}},
  \bibinfo{author}{\bibfnamefont{J.}~\bibnamefont{Furthm\"uller}},
  \bibnamefont{and}
  \bibinfo{author}{\bibfnamefont{F.}~\bibnamefont{Bechstedt}},
  \bibinfo{journal}{Phys. Rev. B} \textbf{\bibinfo{volume}{70}},
  \bibinfo{pages}{045305} (\bibinfo{year}{2004}{\natexlab{a}}).

\bibitem[{\citenamefont{Stekolnikov
  et~al.}(2004{\natexlab{b}})\citenamefont{Stekolnikov, Furthm\"uller, and
  Bechstedt}}]{Stekolnikov04b}
\bibinfo{author}{\bibfnamefont{A.~A.} \bibnamefont{Stekolnikov}},
  \bibinfo{author}{\bibfnamefont{J.}~\bibnamefont{Furthm\"uller}},
  \bibnamefont{and}
  \bibinfo{author}{\bibfnamefont{F.}~\bibnamefont{Bechstedt}},
  \bibinfo{journal}{Phys. Rev. Lett} \textbf{\bibinfo{volume}{93}},
  \bibinfo{pages}{136104} (\bibinfo{year}{2004}{\natexlab{b}}).

\bibitem[{\citenamefont{Stekolnikov et~al.}(2002)\citenamefont{Stekolnikov,
  Furthm\"uller, and Bechstedt}}]{Stekolnikov02}
\bibinfo{author}{\bibfnamefont{A.~A.} \bibnamefont{Stekolnikov}},
  \bibinfo{author}{\bibfnamefont{J.}~\bibnamefont{Furthm\"uller}},
  \bibnamefont{and}
  \bibinfo{author}{\bibfnamefont{F.}~\bibnamefont{Bechstedt}},
  \bibinfo{journal}{Phys. Rev. B} \textbf{\bibinfo{volume}{65}},
  \bibinfo{pages}{115318} (\bibinfo{year}{2002}).

\bibitem[{\citenamefont{Stekolnikov
  et~al.}(2003{\natexlab{b}})\citenamefont{Stekolnikov, Furthm\"uller, and
  Bechstedt}}]{Stekolnikov03b}
\bibinfo{author}{\bibfnamefont{A.~A.} \bibnamefont{Stekolnikov}},
  \bibinfo{author}{\bibfnamefont{J.}~\bibnamefont{Furthm\"uller}},
  \bibnamefont{and}
  \bibinfo{author}{\bibfnamefont{F.}~\bibnamefont{Bechstedt}},
  \bibinfo{journal}{Phys. Rev. B} \textbf{\bibinfo{volume}{68}},
  \bibinfo{pages}{205306} (\bibinfo{year}{2003}{\natexlab{b}}).

\bibitem[{\citenamefont{Hohenberg and Kohn}(1964)}]{Hoh64}
\bibinfo{author}{\bibfnamefont{P.}~\bibnamefont{Hohenberg}} \bibnamefont{and}
  \bibinfo{author}{\bibfnamefont{W.}~\bibnamefont{Kohn}},
  \bibinfo{journal}{Phys.~Rev.~} \textbf{\bibinfo{volume}{136 B}},
  \bibinfo{pages}{864} (\bibinfo{year}{1964}).

\bibitem[{\citenamefont{Kohn and Sham}(1965)}]{Koh65}
\bibinfo{author}{\bibfnamefont{W.}~\bibnamefont{Kohn}} \bibnamefont{and}
  \bibinfo{author}{\bibfnamefont{L.~J.} \bibnamefont{Sham}},
  \bibinfo{journal}{Phys.~Rev.~} \textbf{\bibinfo{volume}{140 A}},
  \bibinfo{pages}{1133} (\bibinfo{year}{1965}).

\bibitem[{\citenamefont{Perdew and Zunger}(1981)}]{Per81}
\bibinfo{author}{\bibfnamefont{J.~P.} \bibnamefont{Perdew}} \bibnamefont{and}
  \bibinfo{author}{\bibfnamefont{A.}~\bibnamefont{Zunger}},
  \bibinfo{journal}{Phys.~Rev.~B} \textbf{\bibinfo{volume}{23}},
  \bibinfo{pages}{5048} (\bibinfo{year}{1981}).

\bibitem[{\citenamefont{Ceperley and Alder}(1980)}]{Cep80}
\bibinfo{author}{\bibfnamefont{D.~M.} \bibnamefont{Ceperley}} \bibnamefont{and}
  \bibinfo{author}{\bibfnamefont{B.~J.} \bibnamefont{Alder}},
  \bibinfo{journal}{Phys.~Rev.~Lett.~} \textbf{\bibinfo{volume}{45}},
  \bibinfo{pages}{566} (\bibinfo{year}{1980}).

\bibitem[{\citenamefont{{\tt http://www.pwscf.org}}()}]{PWscf}
\bibinfo{author}{\bibnamefont{{\tt http://www.pwscf.org}}}.

\bibitem[{\citenamefont{von Barth and Car}(unpublished)}]{CarUn}
\bibinfo{author}{\bibfnamefont{U.}~\bibnamefont{von Barth}} \bibnamefont{and}
  \bibinfo{author}{\bibfnamefont{R.}~\bibnamefont{Car}}
  (\bibinfo{year}{unpublished}).

\bibitem[{\citenamefont{DalCorso et~al.}(1993)\citenamefont{DalCorso, Baroni,
  Resta, and de~Gironcoli}}]{Cor93}
\bibinfo{author}{\bibfnamefont{A.}~\bibnamefont{DalCorso}},
  \bibinfo{author}{\bibfnamefont{S.}~\bibnamefont{Baroni}},
  \bibinfo{author}{\bibfnamefont{R.}~\bibnamefont{Resta}}, \bibnamefont{and}
  \bibinfo{author}{\bibfnamefont{S.}~\bibnamefont{de~Gironcoli}},
  \bibinfo{journal}{Phys.~Rev.~B} \textbf{\bibinfo{volume}{47}},
  \bibinfo{pages}{3588} (\bibinfo{year}{1993}).

\bibitem[{\citenamefont{Monkhorst and Pack}(1976)}]{Mon76}
\bibinfo{author}{\bibfnamefont{H.~J.} \bibnamefont{Monkhorst}}
  \bibnamefont{and} \bibinfo{author}{\bibfnamefont{J.~D.} \bibnamefont{Pack}},
  \bibinfo{journal}{Phys.~Rev.~B} \textbf{\bibinfo{volume}{13}},
  \bibinfo{pages}{5188} (\bibinfo{year}{1976}).

\bibitem[{\citenamefont{Farid and Godby}(1991)}]{Farid91}
\bibinfo{author}{\bibfnamefont{B.}~\bibnamefont{Farid}} \bibnamefont{and}
  \bibinfo{author}{\bibfnamefont{R.~W.} \bibnamefont{Godby}},
  \bibinfo{journal}{Phys. Rev. B} \textbf{\bibinfo{volume}{43}},
  \bibinfo{pages}{14248} (\bibinfo{year}{1991}).

\bibitem[{\citenamefont{Harrison}(1980)}]{Harrison80}
\bibinfo{author}{\bibfnamefont{W.~A.} \bibnamefont{Harrison}},
  \emph{\bibinfo{title}{Electronic Structure and Properties of Solids}}
  (\bibinfo{publisher}{Freeman}, \bibinfo{address}{San Francisco},
  \bibinfo{year}{1980}).

\bibitem[{\citenamefont{Dabrowski and M\"{u}ssig}(2000)}]{Dabrowski00}
\bibinfo{author}{\bibfnamefont{J.}~\bibnamefont{Dabrowski}} \bibnamefont{and}
  \bibinfo{author}{\bibfnamefont{H.~J.} \bibnamefont{M\"{u}ssig}},
  \emph{\bibinfo{title}{Silicon Surfaces and Formation of Interfaces}}
  (\bibinfo{publisher}{World Scientific}, \bibinfo{address}{Singapore},
  \bibinfo{year}{2000}).

\bibitem[{\citenamefont{Battaglia et~al.}(2009)\citenamefont{Battaglia,
  Ga\'{a}l-Nagy, Monney, Didiot, Schwier, Garnier, Onida, and
  Aebi}}]{Battaglia08b}
\bibinfo{author}{\bibfnamefont{C.}~\bibnamefont{Battaglia}},
  \bibinfo{author}{\bibfnamefont{K.}~\bibnamefont{Ga\'{a}l-Nagy}},
  \bibinfo{author}{\bibfnamefont{C.}~\bibnamefont{Monney}},
  \bibinfo{author}{\bibfnamefont{C.}~\bibnamefont{Didiot}},
  \bibinfo{author}{\bibfnamefont{E.~F.} \bibnamefont{Schwier}},
  \bibinfo{author}{\bibfnamefont{M.~G.} \bibnamefont{Garnier}},
  \bibinfo{author}{\bibfnamefont{G.}~\bibnamefont{Onida}}, \bibnamefont{and}
  \bibinfo{author}{\bibfnamefont{P.}~\bibnamefont{Aebi}}, \bibinfo{journal}{J.
  Phys.: Condens. Matter} \textbf{\bibinfo{volume}{21}},
  \bibinfo{pages}{013001} (\bibinfo{year}{2009}).

\bibitem[{\citenamefont{Ga\'al-Nagy and Onida}(2007)}]{GaalNagy07}
\bibinfo{author}{\bibfnamefont{K.}~\bibnamefont{Ga\'al-Nagy}} \bibnamefont{and}
  \bibinfo{author}{\bibfnamefont{G.}~\bibnamefont{Onida}},
  \bibinfo{journal}{Phys. Rev. B} \textbf{\bibinfo{volume}{75}},
  \bibinfo{pages}{155331} (\bibinfo{year}{2007}).

\end{thebibliography}

\end{document}